\documentclass[%
 reprint,
 amsmath,amssymb,
 aps,
]{revtex4-2}

\usepackage{graphicx}
\usepackage{dcolumn}
\usepackage{bm}

\begin{document}

\preprint{APS/123-QED}

\title{Orders of Magnitude Improved Cyclotron-Mode Cooling for Non-Destructive Spin Quantum Transition Spectroscopy with Single Trapped Antiprotons}

\author{B. M. Latacz$^{1,2}$}
\author{M. Fleck$^{1,3}$}
\author{J. I. J{\"a}ger$^{1,2,4}$}
\author{G. Umbrazunas$^{1,5}$}
\author{B. P. Arndt$^{1,4,6}$}
\author{S. R. Erlewein$^{1,4}$}
\author{E. J. Wursten$^{1}$}
\author{J. A. Devlin$^{1,2}$}
\author{P. Micke$^{1,2,4}$}
\author{F. Abbass$^{7}$}
\author{D. Schweitzer$^{7}$}
\author{M. Wiesinger$^{4}$}
\author{C. Will$^{4}$}
\author{H. Yildiz$^{7}$}
\author{K. Blaum$^{4}$}
\author{Y. Matsuda$^{3}$}
\author{A. Mooser$^{4}$}
\author{C. Ospelkaus$^{8,9}$}
\author{C. Smorra$^{1,7}$}
\author{A. Soter$^{5}$}
\author{W. Quint$^{6}$}
\author{J. Walz$^{7,10}$}
\author{Y. Yamazaki$^{1}$}
\author{S. Ulmer$^{1,11}$}
 
\affiliation{$^1$RIKEN, Ulmer Fundamental Symmetries Laboratory, 2-1 Hirosawa, Wako, Saitama, 351-0198, Japan\\
$^2$CERN, Esplanade des Particules 1, 1217 Meyrin, Switzerland\\
$^3$Graduate School of Arts and Sciences, University of Tokyo, 3-8-1 Komaba, Meguro, Tokyo 153-0041, Japan\\
$^4$Max-Planck-Institut f{\"u}r Kernphysik, Saupfercheckweg 1, D-69117, Heidelberg, Germany\\
$^5$Eidgen{\"o}ssisch Technische Hochschule Z{\"u}rich, R{\"a}mistrasse 101, 8092 Z{\"u}rich, Switzerland\\
$^6$GSI-Helmholtzzentrum f{\"u}r Schwerionenforschung GmbH, Planckstraße 1, D-64291 Darmstadt, Germany\\
$^7$Institut f{\"u}r Physik, Johannes Gutenberg-Universit{\"a}t, Staudinger Weg 7, D-55099 Mainz, Germany\\
$^8$Institut f{\"u}r Quantenoptik, Leibniz Universit{\"a}t, Welfengarten 1, D-30167 Hannover, Germany\\
$^9$Physikalisch-Technische Bundesanstalt, Bundesallee 100, D-38116 Braunschweig, Germany\\
$^{10}$Helmholtz-Institut Mainz, Johannes Gutenberg-Universit{\"a}t, Staudingerweg 18, D-55128 Mainz, Germany\\
$^{11}$Heinrich Heine University, D{\"u}sseldorf, Univerist{\"a}tsstrasse 1, D-40225 D{\"u}sseldorf, Germany}%

\collaboration{BASE Collaboration}

\date{\today}

\begin{abstract}
We demonstrate efficient sub-thermal cooling of the modified cyclotron mode of a single trapped antiproton and reach particle temperatures $T_+=E_+/k_\text{B}$ below $200\,$mK in preparation times shorter than $500\,$s. This corresponds to the fastest resistive single-particle cyclotron cooling to sub-thermal temperatures ever demonstrated. By cooling trapped particles to such low energies, we demonstrate the detection of antiproton spin transitions with an error-rate $<0.000025$, more than three orders of magnitude better than in previous best experiments. This method will have enormous impact on multi-Penning-trap experiments that measure magnetic moments with single nuclear spins for tests of matter/antimatter symmetry, high-precision mass-spectrometry, and measurements of electron $g$-factors bound to highly-charged ions that test quantum electrodynamics. 
\end{abstract}

\maketitle


Experiments conducted with single particles in Penning traps play a crucial role in achieving ultra-high precision measurements of masses \cite{myers2019high}, magnetic moments \cite{schneider2017double}, and fundamental constants \cite{hanneke2008new}. Moreover, they provide stringent tests for the fundamental symmetries of the Standard Model of particle physics. Penning traps have been instrumental in performing the most precise direct tests of charge-parity-time  reversal (CPT) invariance in both the lepton sector \cite{van1987new} and the baryon sector \cite{borchert202216}. These traps also enable ultra-precise measurements that test quantum electrodynamics \cite{Fan_PhysRevLett.130.071801, blaum2020perspectives} and contribute to searches for exotic physics  \cite{eliseev2015direct}. 
General limitations in precision Penning-trap studies are caused by magnetic field $B_0$ and electrostatic trap imperfections that lead to particle-energy dependent scaling of the measured cyclotron frequency $\nu_c=(q B_0)/(2\pi m)$ and the spin precession frequency $\nu_L$  \cite{borchert202216, ketter2014classical},   $(q/m)$ is the charge-to-mass ratio of the trapped particle. These energy-dependent frequency shifts impose systematic shifts and uncertainties in the determination of fundamental constants, and limit fractional accuracy.  Furthermore, many experiments that employ coherent techniques for measuring cyclotron frequencies of individual trapped particles \cite{heisse2017high, sturm2014high} face limitations due to cyclotron energy scatter, which scales proportionally to the thermal phase-space volume of the initial energy distribution \cite{borchert2021challenging}. In the case of measuring nuclear magnetic moments such as that of the proton \cite{schneider2017double}, the antiproton \cite{smorra2017parts}, or of $^3\text{He}^{2+}$ \cite{schneider2022direct}, the fidelity of spin state detection in single-particle quantum transition spectroscopy experiments is constrained by the energy $E_+$ in the cyclotron mode, while employing incoherent quantum-spectroscopy techniques \cite{smorra2017parts}.  This necessitates the development of efficient cooling techniques that can reliably achieve particle energies lower than those attained by the commonly used resistive cooling systems in these experiments. \\
In this manuscript we report on the implementation of a sub-thermal cooling device, that consists of two stacked Penning traps, a cooling trap (CT) and an analysis trap (AT), and its attached particle manipulation electronics. The CT is equipped with a high-Q RLC circuit resonant at $\approx 28.6\,$MHz, that acts at its resonance as efficient cooling resistor \cite{ulmer2013cryogenic} and features high particle-to-detector coupling.  The AT has a strong magnetic inhomogeneity superimposed \cite{ulmer2011observation}, and allows to determine modified cyclotron energies $E_+$ with $0.86\,\mu$eV resolution in averaging times of  $\approx10\,$s. 
We successfully demonstrate the preparation of a single trapped antiproton with a cyclotron temperature $T_+=E_+/k_B$ below 200$\,$mK, which is suitable for quantum-spectroscopy experiments with single antiproton spins with an error-rate below 0.0005$\,\%$ \cite{smorra2017observation}, more than 1000 times better than in previous best experiment. The typical preparation time for achieving these conditions is approximately 500 seconds, which is more than 80 times faster compared to our previous experiments \cite{smorra2017parts}. This achievement represents the most rapid cooling of the modified cyclotron mode for a single trapped antiproton observed to date, and has enormuos impact on high precision comparisons on then fundamental properties of protons and antiprotons, as well as heavier nuclei, such as $^3\text{He}^{2+}$.
\\
Our experimental setup involves a superconducting magnet with a horizontal bore, which operates at a magnetic field strength of $B_0=1.945\,$T. Inside the magnet bore, we have positioned our cryogenic multi-Penning trap system, housed in a hermetically sealed vacuum chamber, in which pressures below $10^{-18}\,$mbar are achieved \cite{sellner2017improved}. 
\begin{figure}[htb]
      \centerline{\includegraphics[width=8.5cm,keepaspectratio]{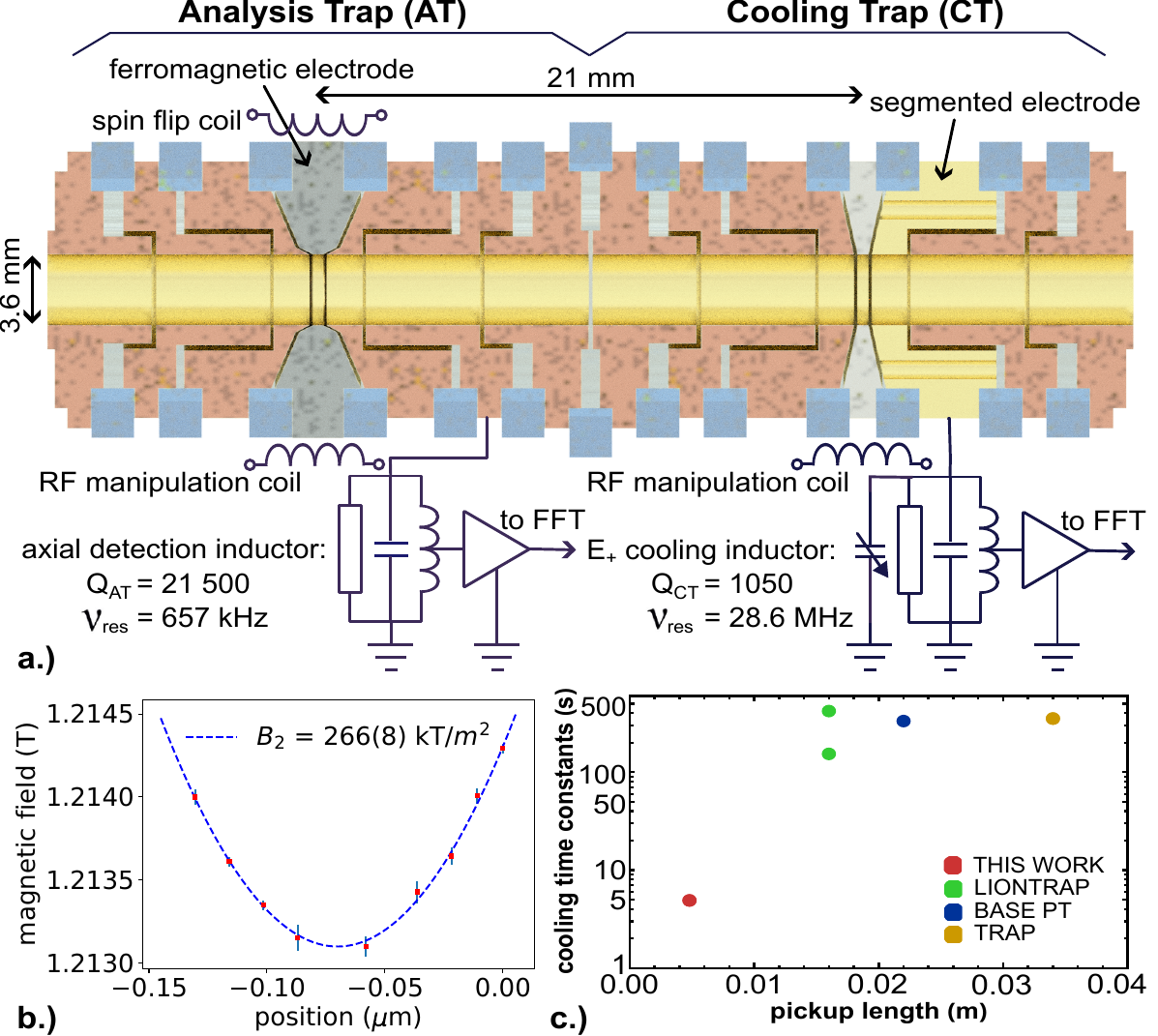}}
     \caption{a.) Double trap system, consisting of a cooling trap (CT) and an analysis trap (AT), with a superimposed magnetic bottle with a strength of $B_{2,\text{AT}}=266(8)\,$kT/m$^2$. The CT is equipped with a resonant cooling resistor to thermalize the modified cyclotron mode. To the end-cap electrode of the AT an $Q\approx21\,500$ superconducting detection system at 657$\,$kHz is connected. b.) magnetic field in the center of the AT.  c.) Comparison of cooling time constants for different trap experiments. Our experiment is operated at $\tau_{+,\text{CT}}\approx4.5\,$s, more than 10 times faster than other state-of-the-art trap experiments.}
     \label{Fig.1}
\end{figure}
For the work reported here, we have extended our multi-Penning trap stack \cite{Smorra2015BASEExperiment} by the CT, and use the AT/CT two trap system shown in Fig$.\,$1 a.). Both traps are in cylindrical five-electrode geometry \cite{gabrielse1989open} with an inner electrode diameter of $3.6\,$mm. A carefully rf-shielded helical resonant cooling resistor \cite{ulmer2013cryogenic, macalpine1959coaxial} is connected to a radially segmented correction electrode of the CT. This device is cooled to environmental temperature of $\approx 5\,$K, and has a varactor-based resonance frequency tuning bandwidth of 4.5$\,$MHz around a center frequency of 26.8$\,$MHz. To suppress stray-noise pickup of electromagnetic interference and to maximize the system inductance $L_\text{CT}$, the cooling resistor is mounted inside the trap vacuum chamber. In thermal equilibrium, the quality factor of the resonant cooling resistor is at $Q\approx1050(50)$, leading with $L\approx1.78(3)\,\mu$H and a resonance frequency of $\nu_{+,\text{CT}}\approx 28.632\,$MHz, to a cooling resistance of $R_{\text{p,CT}}=2\pi\nu_{+,\text{CT}} Q L\approx340\,$k$\Omega$. Once the detector's resonance frequency is tuned to the modified cyclotron mode $\nu_{+,\text{CT}}$ of a single particle in this trap, particle/detector interaction thermalizes the modified cyclotron energy $E_+$ within a correlation time of \cite{wineland1975principles}
\begin{eqnarray}
\tau_{+,\text{CT}}=\frac{m}{R_p}\left(\frac{D_{+,\text{eff}}}{q}\right)^2,
\end{eqnarray}
here $D_{+,\text{eff}}=4.82\,$mm is the effective pickup length of the trap at the chosen electrode geometry. By design we expect $\tau_\text{+,\text{CT},D}=4.5(2)\,$s, corresponding to a more than 10-fold improvement compared to other state-of-the-art trap experiments \cite{schneider2017double,heisse2017high,gabrielse1999precision}, and to a 100-fold reduction compared to the performance of our previous antiproton experiments \cite{smorra2017parts}, see Fig$.\,$\ref{Fig.1} c.). \\
The magnetic field $B_\text{AT}(z)=B_{0,\text{AT}}+B_{2,\text{AT}}z^2$ of the analysis trap has a magnetic bottle with a strength of $B_{2,\text{AT}}=266(8)\,$kT/m$^2$ superimposed, as shown in Fig$.\,$\ref{Fig.1} b.), the background magnetic field of that trap is $B_{0,\text{AT}}=1.212\,$T. The strong $B_{2,\text{AT}}$ is used for the high-resolution determination of the modified cyclotron energy $E_+$, reducing the determination of $E_+$ to a measurement of the axial frequency $\nu_{z,\text{AT}}$. In the strong magnetic inhomogeneity, the orbital magnetic moment $\mu_+=(q/m)(E_+/\omega_+)$ of the modified cyclotron mode is coupled to the axial frequency $\nu_{z,\text{AT}}$, which becomes 
\begin{eqnarray}
\nu_{z,\text{AT}}=\nu_{z, 0,\text{AT}}+\frac{1}{4\pi^2m\nu_{z, 0,\text{AT}}}\frac{B_{2,\text{AT}}}{B_{0,\text{AT}}}E_+,
\end{eqnarray}
where $\nu_{z, 0,\text{AT}}\approx657.92(1)\,$kHz. For the parameters of our trap, the axial frequency $\nu_{z,\text{AT}}$ of a single trapped (anti)proton shifts by $(1/k_\text{B})\cdot(d\nu_{z,\text{AT}}/dT_+)=69.7(5)\,$Hz per $1\,$K in the modified cyclotron mode, corresponding to an $E_+$ energy resolution of 14.7$\,$mK, or 1.2$\,\mu$eV per 1$\,$Hz axial frequency shift. \\   
To determine the axial frequency $\nu_{z,\text{AT}}$ of the trapped proton in the analysis trap, a superconducting detection system with a quality factor of about $Q_\text{AT}\approx21\,500$, a detection inductance $L_\text{AT}\approx1.7\,$mH and a signal-to-noise ratio of 27$\,$dB is used \cite{nagahama2016highly}. Appropriate adjustment of the voltages applied to the trap electrodes tunes the particles to resonance with the axial detector. In thermal equilibrium with the detection system, the particle shorts the thermal noise of the device at the particle's resonance frequency \cite{wineland1975principles}, and the axial frequency $\nu_{z,\text{AT}}$ is determined by fitting a well-understood resonance line to the Fast Fourier Transform (FFT) of the recorded time-transient signal.\\ 
An important factor for the execution of efficient sub-thermal cooling protocols is the optimization of the time that is required to determine $\nu_{z,\text{AT}}$.  To efficiently sample axial frequencies, we subtract detector reference shots with 800$\,$Hz bandwidth from spectra that include particle signatures, and identify the signature created by the particle at its axial frequency using peak threshold detection.  The selected frequency window of 800$\,$Hz covers an equivalent $E_+/k_\text{B}$-temperature range of $\approx12\,$K, sufficient to resolve measured $E_+$ distributions at appropriate resolution. By tuning the trap parameters such that cold particles appear within the 3$\,$dB-width of the detection resonator, within a spectrum averaging time of $\approx10\,$s particles with energies $E_+<440\,$mK can be identified with better than $2\,\sigma$ detection significance.   
\\
For the preparation of a particle with low modified cyclotron energy $E_+$, we first prepare a single particle in the AT, and cool its magnetron mode to an energy of $E_-/k_\text{B}=T_-<7(1)\,$mK, using similar techniques as the ones described in \cite{cornell1990mode,guise2010self, nagahama2017sixfold}.  
\begin{figure}[htb]
      \centerline{\includegraphics[width=8.5cm,keepaspectratio]{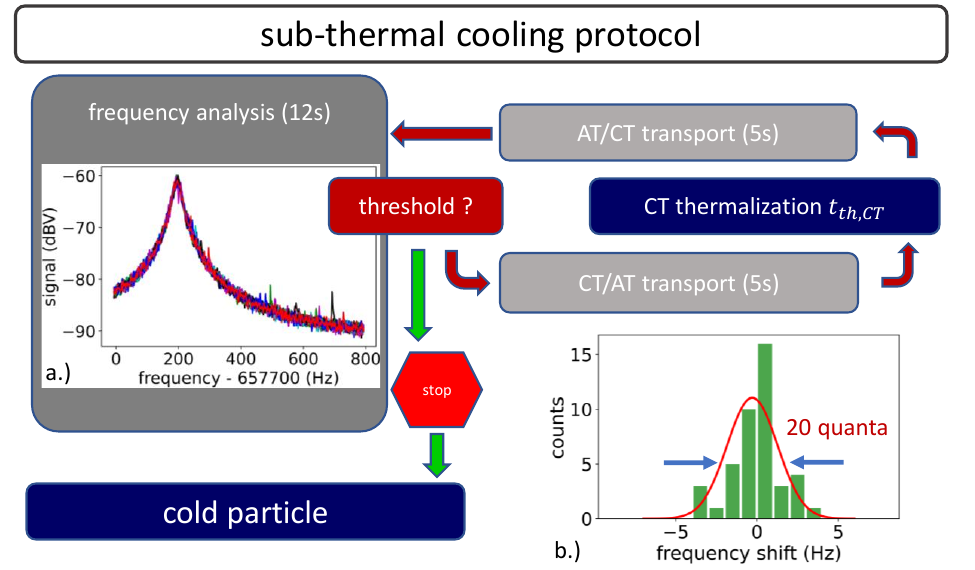}}
     \caption{Flow-chart of the sub-thermal cooling protocol that is described in detail in the text. We cycle frequency measurements, particle transports, and particle thermalization in the CT until a threshold energy $E_{+,\text{th}}$ is reached. For energies $E_+<E_{+,\text{th}}$ the sequence is stopped. a.) illustrates axial resonator spectra, the peak features are signatures of particles at different $E_+$. b.) illustrates the measured trap-to-trap transport scatter, which is for AT$\rightarrow$CT$\rightarrow$AT at 20 cyclotron quanta per attempt. }
     \label{fig:SCT}
\end{figure}
Subsequently, we apply the scheme illustrated in Fig$.\,$\ref{fig:SCT}. First, we  determine the particle's axial frequency by recording a single $12\,$s detector FFT spectrum as explained above. Next, by applying voltage ramps to 10 electrodes that connect the traps, the particle is shuttled from the AT to the CT. One transport takes $4.8\,$s, currently limited by the time constants of filters that connect the voltage supply lines to the trap electrodes. The used transport routines induce a scatter of about 14(2) cyclotron quanta per executed transport protocol, see left lower inset in Fig$.\,$\ref{fig:SCT}. This induces in temperature determinations a fluctuation background of 18$\,$mK ($\approx$1.2$\,$Hz), negligibly small compared to the several 100$\,$Hz fluctuations that need to be resolved. Subsequently, the particle is brought for a time $t_\text{th,CT}$ in contact with the radial thermalization resistor $R_{p,\text{CT}}$ in the CT. Using the varactor \cite{ulmer2013cryogenic}, the resonance frequency of this device is tuned to the CT modified cyclotron frequency, that was earlier determined by single particle cyclotron resonance spectroscopy \cite{nagahama2017sixfold}. Next, the particle is shuttled back to the AT ($4.6\,$s), and its axial frequency is determined again. Following this protocol, one thermalization cycle requires about 22$\,$s of frequency measurement and particle shuttling time, as well as the time $t_\text{th,CT}$ for thermalization of the modified cyclotron mode in the CT. 
\begin{figure}[htb]
      \centerline{\includegraphics[width=8.5cm,keepaspectratio]{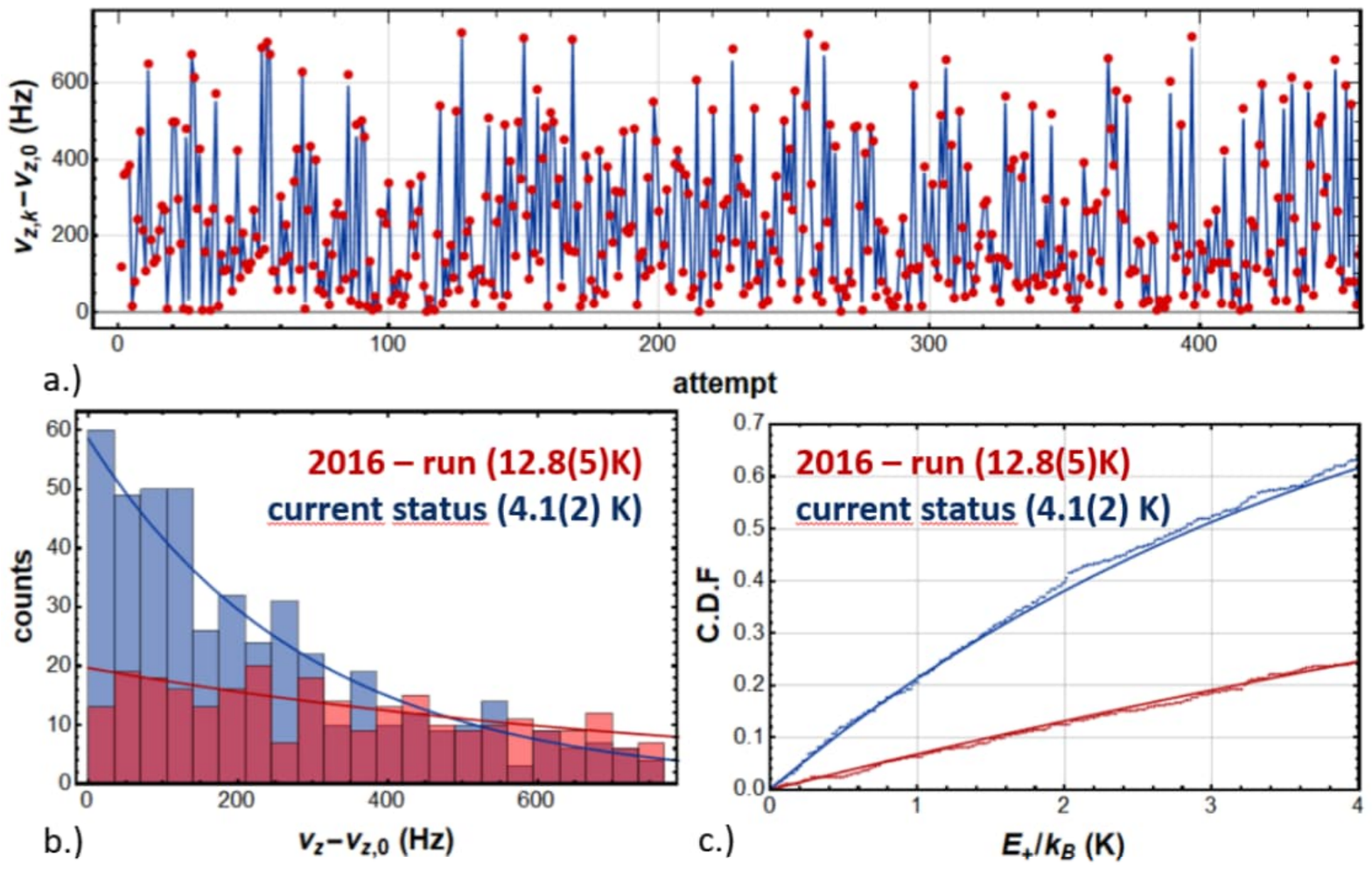}}
     \caption{a.) Frequencies measured in the AT within the sub-thermal cooling procedure. b.) Data shown in a.) plotted to a histogram (blue), and compared to measurements performed with an earlier version of the apparatus c.) Cumulative density functions of the results shown in a.).}
     \label{Fig:CTRES}
\end{figure}
In contact with $R_{p,\text{CT}}$, the particle's modified cyclotron mode is performing a random walk in $E_+$ energy space, once decoupled from the thermal bath, the walk ``freezes" at a modified cyclotron energy $E_{+,\text{CT},k}$, and reaches the AT with a radial orbital magnetic moment $\mu_{+,k}(E_+)=(q_0 E_{+,k})/(2\pi m_p \nu_{+})$, inducing the $E_+$ dependent axial frequency shift $\Delta\nu_{z}(T_+)$.  From experiments where $t_\text{th,CT}=20\,$s was used, we obtain the AT axial frequencies shown in Fig$.\,$\ref{Fig:CTRES} a.). These data represent an $\approx 12\,$K truncated Boltzmann distribution of a weakly-bound one-dimensional thermalized oscillator. To determine the mean temperature of the particle's modified cyclotron mode after thermalization in the CT, and hence the temperature of the thermalization resistor $R_{P,\text{CT}}$, we determine the lowest found frequency, use Eq$.\,$2 and the measured $B_{2,\text{AT}}\approx266(8)\,$kT/m$^2$ to scale the measured frequency shifts $\Delta\nu_{z,k}$ to equivalent absolute temperatures $T_k$. 
Then we determine the number of measured events $N_k(T_{k+1}-T_{k})$ in the temperature interval $\Delta T_+=(T_{+,k+1}-T_{+,k})=0.001\,$K, and evaluate the normalized cumulative density function $1/N_0\cdot\sum_k^{N_0} (N_k(T_{k+1}-T_{k}))$, where $N_0$ is the number of executed thermalization cycles. We fit to the resulting data the thermal cumulative density function $\text{CDF}(T_{+,\text{CT}})=\left((1-\exp\left(-T/T_{+,\text{CT}}\right)\right)$, from which we extract the temperature of the thermalization resistor $T_{+,\text{CT}}=4.3(1)\,$K. Figures \ref{Fig:CTRES} b.) and c.) show results of this data treatment, blue for the current experiment, and red for thermalization in an apparatus without the cooling trap, where $E_+=12.8(5)\,$K was measured \cite{smorra2017parts}. For fully optimized parameters, the accumulation of the data set shown in Fig$.\,$\ref{Fig:CTRES} requires 3.7$\,$h, whereas acquisition of the data set without cooling trap and optimized transport and readout procedures took 55$\,$h. We account the three-fold reduction in the temperature of $R_{P,\text{CT}}$ to the relocation of the detection resistor closer to the trap, added cryogenic shielding and capacitive decoupling of the radio-frequency particle manipulation lines. \\
To investigate and optimize the limits of particle/detector interaction time $t_\text{th}$, we vary $t_\text{th}$ in the CT, and measure the frequency scatter between the determined axial frequencies $\nu_{z,\text{AT},k}$ before the thermalization and $\nu_{z,\text{AT},k+1}$ afterwards. Illustrated in Fig$.\,$\ref{Fig:CT_t} a.), the width of frequency scatter histograms increases linearly with $t_\text{th}$, consistent with Monte-Carlo simulations of the expected dynamics.   
\begin{figure}[htb]
      \centerline{\includegraphics[width=8.8cm,keepaspectratio]{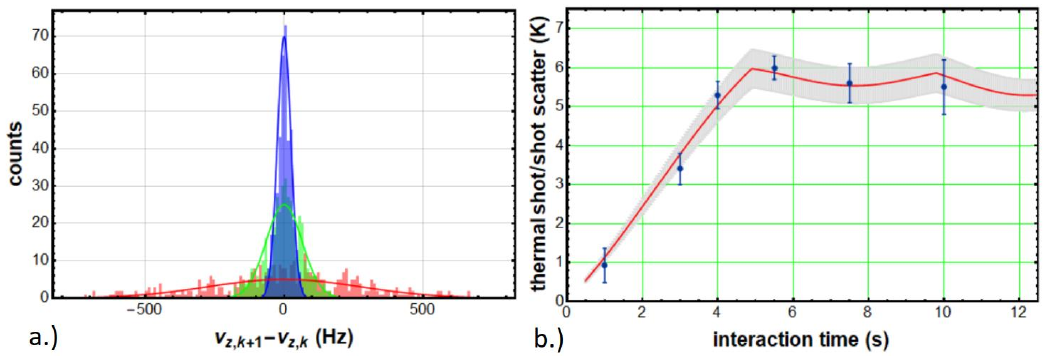}}
     \caption{a.) frequency scatter histograms for different particle/$R_{p,\text{CT}}$ interaction times, blue 0.5$\,$s, green $3\,$s and red $15\,$s.  b.) widths of the histograms shown on the left as a function of widths of histograms shown in a.) as a function of interaction time $t_\text{th}$ in the CT.  }
     \label{Fig:CT_t}
\end{figure}
Once the correlation time $\tau_{+,\text{CT}}$ is reached, the widths of the measured frequency scatter histograms converge to a mean value, weakly $\tau_{+,\text{CT}}$-structured by the correlation time of the thermalization of $E_+$. This behaviour is shown in Fig$.\,$\ref{Fig:CT_t} b.) for the particle parked in the center of the segmented electrode, which is the closest possible distance between particle and detection electrode, and provides the strongest particle-detector coupling. From these measurements we determine $\tau_{+,\text{CT}}=4.7(4)\,$s, within uncertainties in perfect agreement with the theoretically expected $\tau_{+,\text{CT,D}}=4.5(2)\,$s. \\
Applying this CT-based optimized Maxwell-daemon-cooling protocol, we achieve for single trapped antiprotons $E_+/k_B<200\,$mK, at a particle preparation time of $\approx500\,$s, which is the fastest sub-thermal resistive cooling of a single particle in a Penning trap that has ever been demonstrated. \\
We apply our cooling scheme, to demonstrate non-destructive high-fidelity quantum jump spectroscopy with a single antiproton spin using the continuous Stern-Gerlach effect. The strong magnetic bottle $B_{2,\text{AT}}$ of the AT couples the spin magnetic moment $\mu_{\bar{p}}$ to the axial frequency $\nu_{z,\text{AT}}=\nu_{z,0,\text{AT}}\pm\Delta\nu_{z,\text{SF}}/2$, where $\nu_{z,0,\text{AT}}$ is the axial frequency without spin, and $\Delta\nu_{z,\text{SF}}=(\mu_{\bar{p}}B_{2,\text{AT}})/(m_{\bar{p}}\nu_{z,0,\text{AT}}))=173\,$mHz is the axial frequency shift induced by a spin transition.     
\begin{figure}[htb]
      \centerline{\includegraphics[width=8.5cm,keepaspectratio]{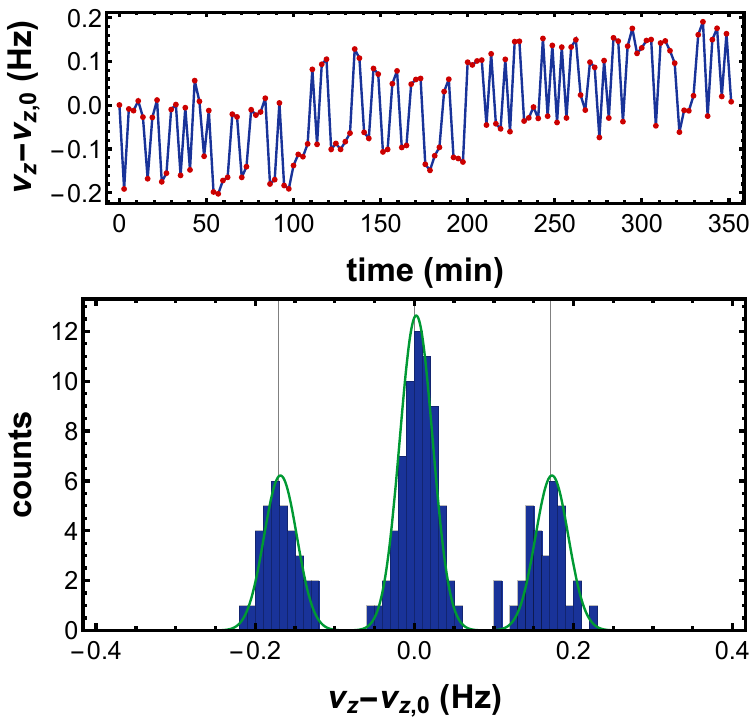}}
     \caption{Top: axial frequency of a single trapped antiproton in the AT, between two subsequent $\nu_{z,\text{AT}}$-measurements a resonant saturating spin flip drive is applied. Bottom: Histogram of measured axial frequency differences $\nu_{z,k+1,\text{AT}}-\nu_{z,k,\text{AT}}$. The left/right part of the histogram represents a $|\downarrow\rangle\rightarrow|\uparrow\rangle$/$|\uparrow\rangle\rightarrow|\downarrow\rangle$ transition, for the central historgam the antiproton spin was not inverted. The rms-widths of the histograms are consistent with the stability of the power supply that biases the trap electrodes of the AT. }
     \label{Fig:SSF}
\end{figure}
Such experiments have been demonstrated with electrons, electrons bound to highly charged ions, as well as protons (p) and antiprotons $\bar{\text{p}}$, being however outstandingly challenging to realize with the p/$\bar{\text{p}}$-system, due to the small magnetic moments $\mu_{p,\bar{p}}$ and the comparably large mass $m_{p,\bar{p}}$. Quantum transitions $\Delta n_+=\pm65\,$mHz in the modified cyclotron mode $E_+$, driven by tiny noise densities on the trap electrodes \cite{borchert2019measurement}, lead to axial frequency fluctuations, and therefore to considerable error-rates in the spin state identification protocols that are applied in Penning trap based measurements of $\mu_{p,\bar{p}}$. The heating rates $d n_+/dt$ scale however $\propto E_+$ \cite{mooser2013resolution}, such that for colder particles the contrast in the spin state identification increases. We apply the sub-thermal cooling protocol and prepare a single antiproton at $E_+/k_B<100\,$mK in the center of the AT. Subsequently, we apply a sequence of axial frequency $\nu_{z,\text{AT}}$ measurements, interleaved with resonant saturated spin-flip drives that incoherently invert the spin state in the AT \cite{brown1986geonium}. Results of these measurements are shown in Fig$.\,$\ref{Fig:SSF}, the upper plot displays the measured frequencies $\nu_{z,\text{AT}}$, the two antiproton spin quantum states $|\downarrow\rangle$  and $|\uparrow\rangle$ can be clearly distinguished.  The lower histogram shows the measured frequency differences $\nu_{z,k+1,\text{AT}}-\nu_{z,k,\text{AT}}$ plotted to a histogram. The sub-histograms at $\Delta\nu_{z,AT}=\pm171(1)\,$mHz represent transitions $|\downarrow\rangle\rightarrow|\uparrow\rangle$/$|\uparrow\rangle\rightarrow|\downarrow\rangle$ respectively, for the central histogram, the antiproton spin was not inverted. The rms-widths of the histograms $\sigma(\nu_{z,\text{AT}})=21(1)\,$mHz are consitent with power-supply and detector-based frequency measurement noise. The error rate $E_S$ - the probability to incorrectly assign an observed frequency shift $\Delta\nu_{z}$ to a spin transition, given a defined detection threshold $\Delta_{\text{TH}}$ - is at that performance of the experiment at a level of $2.3\cdot 10^{-5}$, which corresponds to a $>1500$ times improvement compared to the previous best reported $E_S$ \cite{Smorra2015BASEExperiment}, dominantly thanks to the sub-thermal cooling protocol developed in this work. This also opens perspective towards direct precision measurements of heavier nuclei, such as $^3\text{He}^{2+}$, which is an attractive standard for absolute magnetometry \cite{schneider2017double}. Another application of the cooling presented here are coherent measurements of cyclotron frequencies in high-precision mass spectrometry and measurements of the bound electron $g$-factor  \cite{sturm2014high}. In those experiments, sideband-cooling techniques are applied, that lead to mode-temperature related phase-scatter, which could be considerably suppressed by applying the above cooling trap technique. 
\\

We acknowledge financial support by RIKEN,  the Max-Planck Society, CERN, the European Union (FunI-832848,  STEP-852818), CRC 1227 ``DQ-mat"(DFG 274200144), the Cluster of Excellence ``Quantum Frontiers" (DFG 390837967), the Wolfgang Gentner Program (grant no. 13E18CHA), IMPRS-QD, and the Helmholtz-Gemeinschaft. This work was supported by the Max-Planck, RIKEN, PTB-Center for Time, Constants, and Fundamental Symmetries (C-TCFS).

\bibliography{apssamp}

\end{document}